\documentclass[11pt]{article}
\usepackage{authblk}
\usepackage{caption}
\usepackage[caption=false]{subfig}

\usepackage{amsmath,amsfonts,amsthm,amssymb,graphics,graphicx,mathtools,epstopdf,etoolbox,indentfirst,enumitem,mleftright,xcolor,booktabs,footnote,relsize}

\usepackage[
backend=bibtex,
style=alphabetic,
giveninits=true, 
maxbibnames=99,
minalphanames=3,
date=year
]{biblatex}
\usepackage{hyperref} 
\usepackage{xurl} 
\definecolor{darkmagenta}{rgb}{0.85, 0, 0.45}
\hypersetup{
breaklinks=true,
colorlinks = true,
citecolor= blue,
urlcolor= blue,
linkcolor = darkmagenta
}
\usepackage{doi} 

\renewbibmacro{in:}{}

\AtEveryBibitem{\clearfield{issue}} 

\newbibmacro{string+doiurl}[1]{%
	\iffieldundef{doi}{%
		\iffieldundef{url}{
			#1%
		}{%
			\href{\thefield{url}}{#1}%
		}%
	}{%
		\href{http://dx.doi.org/\thefield{doi}}{#1}%
	}%
}
\DeclareFieldFormat{title}{\usebibmacro{string+doiurl}{\mkbibemph{#1}}}
\DeclareFieldFormat[article,thesis,incollection,inproceedings,manual]{title}%
{\usebibmacro{string+doiurl}
	{#1} 
}
\usepackage{algorithm} 
\usepackage[noend]{algpseudocode}
\floatname{algorithm}{Protocol} 
\algrenewcommand\alglinenumber[1]{\normalsize #1.} 

\newcounter{algsubstate}




\newcommand{\ket}[1]{\left| #1 \right>}
\newcommand{\bra}[1]{\left< #1 \right|}
\newcommand{\ketbra}[2]{\ket{#1} \! \bra{#2}}
\newcommand{\pure}[1]{\ketbra{#1}{#1}}
\newcommand{\binh}{h_2} 

\newcommand{\eps}{\varepsilon}

\newcommand{\pct}{\%} 

\newcommand{\term}[1]{\textbf{#1}}

\newcommand{\qA}{A}
\newcommand{\qB}{B}

\newcommand{\allE}{\mathsf{E}}

\newtheorem*{remark}{Remark}

\theoremstyle{definition} 

\newtheorem{example}{Example}

\newfloat{process}{htbp}{loa}
\floatname{process}{Process}

\begin{document}

\title{\textbf{Memory effects in device-dependent and device-independent cryptography}}

\renewcommand\Affilfont{\itshape\small} 

\author[1]{Ernest Y.-Z.\ Tan}
\affil[1]{Institute for Quantum Computing and Department
of Physics and Astronomy, University of Waterloo, Waterloo, Ontario N2L 3G1, Canada.}

\date{}

\maketitle


\newcommand{\memchann}{\mathcal{M}}
\newcommand{\mA}{\qA'}
\newcommand{\mB}{\qB'}

In~\cite{BCK13}, various {memory attacks} on device-independent (DI) cryptography were discussed, based on the fact that the devices could fully retain memory of their outputs in one protocol instance, then broadcast them in subsequent protocol instances. This is an introductory note to highlight the fact that even without such an extreme form of memory attack, various possible memory effects in device-dependent and device-independent QKD (within a \emph{single} instance of the protocol, i.e.~entirely without considering device-reuse issues) are enough to cause substantial difficulties in applying existing non-IID proof techniques. While such memory effects are typically excluded by assumption in device-dependent cryptography, the points raised in this note highlight that even a slight weakening of this assumption may be sufficient to introduce loopholes in several existing proof techniques. As for the DI case, these points indicate that those proof techniques do not seem to be straightforwardly applicable in the DI setting, unless one explicitly makes the additional assumption that such memory effects are forbidden, which would weaken the sense in which the proof is ``device-independent''. (Various issues discussed here were already previously noted in a brief comment in the work~\cite{TFK+13} on semi-DI protocols. Furthermore, some recently discussed non-IID attacks~\cite{TTB+16,arx_SW23} on particular DIQKD protocols can be viewed as being related to another of the issues we present.)

\section{Background}

Currently, many non-IID proof techniques for device-dependent QKD (such as de Finetti reductions~\cite{rennerthesis,CKR09}, complementarity-based privacy amplification~\cite{Koa09}, and entropic uncertainty relations~\cite{TR11,TL17}), are designed to apply to device behaviours that can be described by the following model. (We shall use the following notation: the registers Alice and Bob measure in the $j^\text{th}$ round of the protocol will be denoted as $\qA_j$ and $\qB_j$ respectively,
$\allE$ denotes Eve's side-information register (across all the rounds), and when allowing for memory effects, $\mA_j, \mB_j$ denotes ``memory registers'' that the devices retain in each round for use in the next round.\footnote{In principle the memory registers $\mA_j, \mB_j$ can be 
embedded into the Hilbert spaces of the pre-measurement registers $\qA_j,\qB_j$;
however, our subsequent discussion is clearer with separate notation for the two spaces.})
\begin{process}
\caption{} 
\label{proc1}
\begin{algorithmic}[1]
\State Initially, Alice and Bob hold a state $\rho$ on the space $\bigotimes_j \mathcal{H}_{\qA_j \qB_j}$, which can be arbitrarily correlated or entangled across all the rounds, and Eve holds some extension of this state on a register $\allE$. 
\State For each round $j$, the devices perform some measurements (as specified by Alice and Bob, e.g.~by supplying some classical inputs that specify the desired measurement basis) on registers $\qA_j$ and $\qB_j$.
\end{algorithmic}
\end{process}

\newpage
However, in practice, QKD protocols are typically implemented in a sequential round-by-round fashion, with the state in each round only being received after the measurement in the previous round as been completed. Therefore, it does not seem unreasonable to suggest that the device behaviour is instead described by the following model:
\begin{savenotes}
\begin{process}
\caption{} 
\label{proc2}
\begin{algorithmic}[1]
\State Initially, none of the registers $\qA_j \qB_j$ contain a state (equivalently, they just contain some arbitrary ``placeholder'' state that will be overwritten in the next steps). 

\State For each round $j$, the following occurs:

\begin{itemize}
\item For $j=1$ (the first round), Eve prepares a joint state on registers $\qA_1 \qB_1 \allE$, then sends $\qA_1 \qB_1$ to Alice/Bob and keeps $\allE$ as side-information. Alice's device performs a measurement on $\qA_1$ (as specified by Alice's input), then generates a state on a memory register $\mA_1$. Analogously, Bob's device measures $\qB_1$ and produces a memory register $\mB_1$. 

\item For $j>1$ (all subsequent rounds), Eve applies a CPTP map on her register $\allE$ to produce some joint state on registers $\qA_j \qB_j \allE$, then sends $\qA_j \qB_j$ to Alice/Bob, keeping $\allE$ as side-information (the state on $\allE$ can be updated in each round). Alice's device applies a ``memory channel'' $\memchann^A_j$ on the registers\footnote{In this description, the memory channel uses the memory register from the preceding round only, not the preceding rounds. In the DI case, this is not a significant restriction since all memory from preceding rounds can be fed to the next memory register. On the other hand, in the device-dependent case, the condition that the measurements (and implicitly, system dimensions) are trusted means that this can be a nontrivial restriction. Hence in principle to be sufficiently general in the device-dependent case, we should allow the memory channels to act on all preceding memory registers; however, even the restricted description here is sufficient to demonstrate challenges for some existing proof techniques.} $\mA_{j-1} \qA_j$ that produces some updated state on $\qA_j$. Then it performs a measurement on $\qA_j$ and generates a state on another memory register $\mA_j$. Analogously, Bob's device applies a memory channel $\memchann^B_j$ that maps the registers $\mB_{j-1} \qB_j$ to some updated state on $\qB_j$, then measures $\qB_j$ and produces another memory register $\mB_j$. 
\end{itemize}

\end{algorithmic}
\end{process}
\end{savenotes}

Under a Process~\ref{proc2} model, it is not obvious whether one can speak in a well-defined sense of some joint state on the space $\bigotimes_j \mathcal{H}_{\qA_j \qB_j}$ that describes the device behaviour, because the state in each round is  ``consumed'' in the process of measuring it and preparing the state in the next round. (One could of course consider the state left on $\bigotimes_j \mathcal{H}_{\qA_j \qB_j}$ and/or $\bigotimes_j \mathcal{H}_{\mA_j \mB_j}$ at the very end of the protocol, but this consists entirely of post-measurement states, and hence it is unclear whether it tells us enough for a security proof.)
This immediately raises a number of questions, which we shall discuss in the following sections --- we also provide a concise summary of the answer below each question:

\begin{itemize}
\item Given that QKD protocols in practice are typically implemented sequentially, are there in fact any implementations where Process~\ref{proc1} is a reasonable model to study in the security proof? 

Condensed \hyperref[sec:justify]{answer}: Yes (by a fairly standard argument), \emph{if} we suppose that the devices are sufficiently trusted to enforce
a particular form of no-memory condition (which is usually implicit in the device-dependent case but less standard for the DI case).

\item Are there any scenarios in Process~\ref{proc2} which cannot be mapped to a Process~\ref{proc1} description? 

Condensed \hyperref[sec:diffs]{answer}: Yes, and we give some simple examples.

\item Do these scenarios pose actual problems in applying existing proof techniques for Process~\ref{proc1} to Process~\ref{proc2}? Exactly what are the obstacles that arise?

Condensed \hyperref[sec:problems]{answer}: 
To some extent, yes, they present concrete problems for some existing proof techniques. 
Naive applications of those techniques to the aforementioned scenarios can result in clearly wrong conclusions, and furthermore there is an issue that using those techniques in practice implicitly requires a form of no-signalling (or counterfactual definiteness) property that is violated in Process~\ref{proc2}. 

\item Even if the proof techniques encounter obstacles, are there any concrete {attacks} that are possible under Process~\ref{proc2} but not Process~\ref{proc1}? 

Condensed \hyperref[sec:attacks]{answer}: To some extent --- we describe a contrived QKD protocol which is secure in a Process~\ref{proc1} scenario, and completely insecure in a Process~\ref{proc2} scenario. However, for most existing QKD protocols it seems unlikely that they genuinely become insecure in a Process~\ref{proc2} scenario, because of the next point.

\item Given the above, are there any existing proof techniques that apply to Process~\ref{proc2}? 

Condensed \hyperref[sec:valid]{answer}: Yes, such as entropy accumulation, quantum probability estimation, and parallel-repetition proofs. However, there are a few restrictions in each of these techniques, which we elaborate on.
\end{itemize}

\section{Justifying Process~\ref{proc1}}
\label{sec:justify}

Suppose we assume that in a protocol, the behaviour of the devices follows Process~\ref{proc2}, except with the additional constraint that the memory registers $\mA_j, \mB_j$ are all trivial (or essentially equivalently, that the memory channels $\memchann^A_j, \memchann^B_j$ completely ignore the registers $\mA_{j-1}, \mB_{j-1}$ in their input). In that case, the action of the memory channels can be absorbed into the state preparation process by Eve in each round, and we shall proceed under that model (i.e.~we now ignore the memory channels $\memchann^A_j, \memchann^B_j$, taking them to act as identity channels). Now observe that under this model, Alice and Bob's measurements in round $j$ commute with Eve's state preparation process in all the subsequent rounds (because the round-$j$ measurements act only on registers $\qA_j \qB_j$ and produce no registers that are used in subsequent rounds). This implies that we can defer all the measurements to the end, i.e.~we first implement Eve's preparations of the states in all the registers $\qA_j \qB_j$, then implement the measurements in all the rounds. This exactly matches a Process~\ref{proc1} description; furthermore, it is fairly straightforward to reverse this correspondence to see that every Process~\ref{proc1} description can also be implemented using such a process. 

This line of reasoning highlights the fact that in order to straightforwardly apply the proof techniques for a Process~\ref{proc1} model in device-dependent QKD, we are relying on the notion that the trusted-device assumption implicitly includes a constraint that there are no memory registers $\mA_j, \mB_j$ feeding nontrivial information across the rounds. 
This analysis also suggests the fundamental obstacle in generically mapping 
a Process~\ref{proc2} description to Process~\ref{proc1} --- when the memory registers/channels are nontrivial, the round-$j$ measurements do not necessarily commute with the state preparation process, because the states on registers $\qA_j \qB_j$ depend on the memory registers produced after the measurement in the previous round. We now proceed to give concrete examples which show that such a mapping cannot exist in general.

\section{Highlighting the differences}
\label{sec:diffs}

In the DI context, an example of a Process~\ref{proc2} model that cannot be mapped to Process~\ref{proc1} can be immediately obtained by the following observation: under a Process~\ref{proc1} model, the correlations produced by the devices cannot be \emph{signalling} across the rounds, because the measurements are acting on a tensor product of different Hilbert spaces for each round. 
Therefore, a simple example of a Process~\ref{proc2} model with no Process~\ref{proc1} description would be the following:
\begin{example}
In the first round, the devices produce a uniformly random output for any input. In all subsequent rounds, the device output is simply the input from the previous round (taking the inputs and outputs to have the same set of possible values, which is the case for e.g.~the CHSH game).
\end{example} 
\noindent The resulting correlations are manifestly signalling across the rounds, and hence cannot be reproduced by \emph{any} Process~\ref{proc1} model.

This example, however, does not seem to have any features that clearly showcase difficulties in applying Process~\ref{proc1} proof techniques to a Process~\ref{proc2} model. We hence present another scenario, designed to be relevant even in the device-dependent case, in particular for the analysis of QKD protocols with similar structure to entanglement-based BB84. Suppose that Alice and Bob have devices that in each round can be used to perform either an $X$ or $Z$ measurement, and we work under the assumption that these measurements are trusted (which implicitly also imposes that all the registers $\qA_j \qB_j$ are qubit registers). However, we allow the Process~\ref{proc2} feature of retaining memory after each round. With this in mind, consider the following example:
\begin{example}
\label{ex:seqXZ}
In the first round, each device receives half of the Bell state $\ket{\Phi^+}$, and measures it in the chosen basis. Subsequently, however, the device simply retains the post-measurement state after each round and measures it in the next round (in the chosen basis), without receiving any new states at all. (Here we take the trusted measurements to be ``genuinely projective'' even in the quantum-instrument description, i.e.~if for instance Alice measures $Z$ and gets the outcome $0$, the post-measurement state is the $Z$-basis eigenstate $\ket{0}$, which is then retained and measured in the next round.)
\end{example}

Such a device behaviour has a slightly curious property: in the event that Alice and Bob measure $X$ in all rounds, then the outputs in \emph{all} the rounds are the same (either they are all $0$ or all $1$, with $50\pct$ probability each). Similarly, if they choose to measure $Z$ in all rounds, then all outputs are the same as well. It is not obvious whether such correlations can be reproduced using a Process~\ref{proc1} model --- for instance, observe that if the devices had instead shared the state $\ket{\Phi^+}^{\otimes n}$ at the start and performed $X$/$Z$ measurements on the individual qubits, the outcomes would still be perfectly correlated between Alice and Bob, but instead completely uncorrelated across different rounds, hence differing from the above example. It turns out that it is in fact indeed impossible\footnote{This impossibility claim is restricted to the case where the devices are performing trusted $X$ and $Z$ measurements. In a DI context where each party in each round has two possible inputs corresponding to some untrusted measurements, it is possible to trivially produce such correlations using a Process~\ref{proc1} model even without entanglement --- since the two untrusted measurements do not have to be different from each other, the devices can simply share the state $(\pure{00}^{\otimes n} + \pure{11}^{\otimes n})/2$ and measure everything in the $Z$ basis regardless of the input.} to achieve such correlations via Process~\ref{proc1} --- this will be implied by our discussion in the following section, where we eventually observe that if this were possible, it would violate the security guarantees of some proof techniques that apply to Process~\ref{proc1}.

\section{Obstacles to some proof techniques}
\label{sec:problems}

In this section, we focus specifically on discussing proof techniques in the following categories: de Finetti reductions (including the postselection technique)~\cite{rennerthesis,CKR09}, complementarity-based privacy amplification~\cite{Koa09}, and one-shot entropic uncertainty relations~\cite{TR11} (for brevity, we shall abbreviate these as dF, cPA, and EUR respectively). These proof techniques were originally intended for Process~\ref{proc1}, and we shall discuss the obstacles they face when considering Process~\ref{proc2}.

The first category (dF techniques) faces an immediate obstruction in that they require some form of permutation symmetry in the quantum state across the rounds (or slightly less strictly, some permutation symmetry of the protocol structure). In a general Process~\ref{proc2} scenario, as discussed at the beginning, it is unclear whether there is even any useful well-defined joint state on the space $\bigotimes_j \mathcal{H}_{\qA_j \qB_j}$ that we can speak of as being permutation-symmetric. (One crude attempt would be to consider the pre-measurement quantum state on $\qA_j \qB_j$ for each $j$ and then take the tensor product of all such states; however, this is clearly not a useful state to consider since it does not even have any correlations across rounds. As we shall discuss at the end of this section, 
there seem to be some rather fundamental difficulties in finding a ``useful'' state to consider.)
As for the protocol structure,
it is also unclear whether it can be validly said to have a permutation symmetry, since there is a fundamental time-ordering structure on the rounds --- of course, one could randomly permute the classical data across the rounds, but this might not correspond to any useful operation on the (possibly ill-defined) pre-measurement quantum state across the rounds. (Still, it would at least result in a permutation-symmetric nonlocal distribution, and there is some ongoing work~\cite{AR15,arx_JT21} on finding DIQKD security proofs based only on this property. However, various obstacles remain to be resolved in those approaches.)

As for the other two categories, we first give a very brief informal description of their claims. We focus on BB84-like situations in which all registers are qubits and in each round, the honest parties can choose to perform (trusted) $X$ or $Z$ measurements (more general versions of these techniques can apply to more general notions of ``conjugate measurements'', but we do not discuss them here). Qualitatively, the cPA technique states the following: in a Process~\ref{proc1} scenario, suppose we are given a guarantee that \emph{if} all the registers were measured in the $X$ basis, the fraction of rounds in which Alice and Bob get different outcomes (this fraction is called the \term{phase error rate}) would with high probability be less than some value $\delta_\mathrm{ph}$. (We briefly put aside the question of how to obtain such a guarantee, to be discussed later.) Then if instead Alice measures all her rounds in the $Z$ basis, there exists a privacy amplification procedure she can apply on the resulting outcomes, which produces an $\eps$-secret key of length approximately $\sim\! (1-\binh(\delta_\mathrm{ph}))n$, up to some small $\eps$-dependent corrections. As for EURs, roughly speaking they offer a similar overall statement, but the approach is somewhat different: the smoothed max-entropy (conditioned on Bob) of Alice's $X$-basis outcomes is used to bound the smoothed min-entropy (conditioned on Eve) of Alice's $Z$-basis outcomes\footnote{Technically, a later version of the EUR~\cite{TL17} does not require considering two separate states produced by performing $X$ or $Z$ measurements, instead directly bounding the smooth min-entropy of a state produced by randomly measuring in either basis in each round. However, the theorem is still stated in the context of a Process~\ref{proc1} scenario, in any case.}, then the Leftover Hashing Lemma is invoked.\footnote{In this discussion, we have ignored the question of error correction between Alice and Bob, focusing only on the length of $\eps$-secret key that can be produced (note that this only involves Alice). When error correction is considered (to produce a \emph{shared} secret key), the length of generated key is reduced according to the efficiency of the error-correction protocol --- for the EUR approach, one simply applies a chain rule; for the cPA approach the arguments are somewhat different (e.g.~\cite{Koa09} simply uses a preshared key) and we shall not go into it here.}

In their usual form, the cPA/EUR techniques are described in terms of a joint state across $\bigotimes_j \mathcal{H}_{\qA_j \qB_j}$, in the sense of a Process~\ref{proc1} model. Hence it is again unclear whether they directly apply to Process~\ref{proc2} models in their stated form. Still, one could consider the question of whether they still apply in some sense to Process~\ref{proc2} models. To answer this, we now discuss how ``naive'' interpretations of these theorem statements do not seem to hold in any sense for Process~\ref{proc2} models.

\subsection{The naive approach and the resulting issues}

With the above in mind, we see that Example~\ref{ex:seqXZ} in the previous section poses a challenge for straightforward applications of these techniques. Recall that if Alice and Bob choose to measure $X$ in all rounds, these devices produce perfectly correlated outputs. Hence in a naive application of the cPA or EUR techniques, one might claim that the phase error rate is guaranteed to be zero, and hence approximately $\sim\! n$ bits of secret key could be produced from Alice's outputs if she measured $Z$ in all rounds. However, for these devices, the outputs when $Z$ is measured in all rounds clearly only have $1$ bit of entropy across all the rounds, and hence this conclusion would be completely wrong. (Another perspective on this is that only one ebit of entanglement was used to generate these correlations, which implies (via upper bounds on secret key rate, such as 
intrinsic information~\cite{MW99}) that at most one shared secret key bit can be distilled using public classical communication.) 

Returning to a point mentioned in the previous section, this observation also serves as a proof that the correlations produced in Example~\ref{ex:seqXZ} can never be reproduced by a Process~\ref{proc1} description (with trusted $X$/$Z$ measurements). This is because the correlations produced by the latter would have to obey the security guarantees of these proof techniques, and we have just shown that the Example~\ref{ex:seqXZ} correlations severely violate these guarantees.

In fact, even apart from the misleading conclusion, there is a more subtle issue that technically prevents the above argument from ``getting off the ground'' in the first place when trying to use it in a security proof of an actual protocol.\footnote{This issue has already been briefly noted in~\cite{TFK+13}, which instead derived a monogamy-of-entanglement property that can accommodate a scenario where one party's device is allowed to have memory, but still requires the other party's device to be trusted and memoryless.} Namely, note that it begins with some observations about the correlations that would be produced if $X$ were measured in all rounds, or in other words an estimate or guarantee of the phase error rate $\delta_\mathrm{ph}$. However, it skipped the question of how Alice and Bob would actually know or certify this in practice --- for instance, in any QKD protocol using the $Z$ basis for key generation, the parties would of course not measure $X$ in all rounds, but rather only some ``test-round subset'' (chosen according to the protocol details). To apply the cPA or EUR techniques, one needs to relate the test-round output distribution in the actual protocol to the hypothetical scenario where everything is measured in the $X$ basis.\footnote{It is perhaps worth highlighting that a foundational result in Bell nonlocality, Fine's theorem~\cite{Fine82}, states that it is \emph{impossible} to construct a joint probability distribution across both the $X$ and $Z$ outcomes simultaneously, so the analysis of the output distributions in the two scenarios has to be handled carefully (e.g.~one cannot use any classical statistical analysis that assumes there is a well-defined probability distribution across the $X$ output strings and $Z$ output strings simultaneously). The approach presented in the next paragraph describes how to validly handle it for Process~\ref{proc1} models.}

When only considering Process~\ref{proc1} models, the core property that addresses the above question is this: given the tensor-product structure, the output distribution from the test-round subset is sufficient to yield some conclusions about the output distribution on all the rounds that \emph{would} have been produced if everything had been measured in the $X$ basis. More precisely (focusing on protocols where all test rounds are measured in the $X$ basis by both parties): in the actual protocol some subset is picked (under some distribution) to be the test rounds, and these rounds are measured in the $X$ basis (while the other rounds may be measured in other bases), with the outputs being used for parameter estimation. In a Process~\ref{proc1} model, the no-signalling condition gives the crucial guarantee that the test-round output distribution produced this way is exactly the same as though \emph{all} the rounds had been measured in the $X$ basis first, and then some subset taken as the test rounds (using the same distribution as before), with their outputs being used in parameter estimation. The latter scenario is where the cPA or EUR techniques can be used to obtain the relevant security guarantees.

In contrast, however, for Process~\ref{proc2} models such as Example~\ref{ex:seqXZ}, the marginal distribution on the test-round subset could change dramatically depending on what is measured in the other rounds, since there are no NS constraints across the rounds. Therefore, it is not clear whether the test-round output distribution observed in the actual protocol tells us anything about the phase error rate (in the usual sense of the cPA/EUR techniques, i.e.~conditioned on measuring all rounds in the $X$ basis). 
A security proof based on the cPA/EUR techniques would have to address this point.

\subsection{Difficulties in workarounds}

Of course, the above attempt to apply those proof techniques may be too naive --- perhaps rather than considering the correlations the \emph{actual} devices would produce if (for instance) all measurements were in the $X$ basis, one could instead try to construct some ``virtual''/``effective'' state across all the rounds (keep in mind that as mentioned previously, no such state ``physically'' exists in a Process~\ref{proc2} description), and discuss the correlations one would have obtained if all the registers of \emph{this} state had been measured in the $X$ basis. There is, of course, considerable freedom in how one might try to construct such a state, and we currently do not know any specific issue that would obstruct all attempts along these lines (furthermore, the list of alternative techniques that \emph{do} work in Sec.~\ref{sec:valid} implies that any such obstruction cannot be ``maximally general'' for all protocols, although it is worth noting that these techniques do not involve constructing such a state). 

However, constructing such a state does not seem particularly straightforward.
Basically, it would seem that we would want it to have the following properties: the correlations it produces in the $X$ basis are related to those observed in the test rounds in the protocol, and similarly the values it produces in the $Z$ basis are related to those actually produced in the generation rounds. Furthermore, the outcomes it would produce in the $X$ basis should be related to those it would produce in the $Z$ basis in a way that allows us to apply the cPA/EUR techniques. However, since in Process~\ref{proc2} models the choice of measurements in earlier rounds can arbitrarily affect the outcomes in later rounds, it seems difficult to construct a state with all these properties simultaneously. 
For instance, if we focus on just the first two rounds of the protocol, while there is indeed some state on registers $\qA_1 \qB_1$ before the round 1 measurement, and some state on registers $\qA_2 \qB_2$ before the round 2 measurement, it is not clear if they can be usefully ``connected'' into a joint state. While one could perhaps consider the conditional states on $\qA_2 \qB_2$ conditioned on the measurement choices in the first round, this encounters the issue that if in an actual realization of the protocol Alice and Bob measured $Z$ in round 1 and $X$ in round 2, there seems to be no way to deduce from this any properties of the outcome distribution (across both rounds) that would be obtained if $X$ had been measured in both rounds, because changing the measurement choice in round 1 from $Z$ to $X$ could cause essentially arbitrary changes in the states for round 2. 

From a foundational perspective, the issue here is basically a lack of counterfactual definiteness --- we cannot speak of outcomes that we would have obtained in a particular round (under either measurement choice) without first specifying all the measurement choices in preceding rounds, since these choices could arbitrarily affect the outcomes after them.
(Note that this is the case even with a purely classical Process~\ref{proc2} model --- unlike the typical scenario in e.g.~Bell nonlocality, the measurement choices in different rounds here can ``directly'' influence each other by keeping a memory variable that tracks the measurement choices in past rounds and uses them to determine outcomes in later rounds.
In some ways, the situation here is more closely related to temporal correlations~\cite{CTM+13}, in which arbitrary correlations are possible even with classical models, as long as memory is allowed.)

Another potential approach would be to find an explicit statement of a version of the cPA/EUR theorems that directly applies to Process~\ref{proc2} models, without constructing some virtual state across all the rounds. However, such a statement would need to avoid the potential difficulties presented by Example~\ref{ex:seqXZ}, and also the issue in the following section. 

\section{Scope for attacks}
\label{sec:attacks}

We now present a rather artificial protocol which is secure in a Process~\ref{proc1} scenario, and completely insecure in a Process~\ref{proc2} scenario, hence displaying a ``maximal'' sort of separation between the two cases.
As in previous sections, suppose the devices can perform trusted $X$ or $Z$ measurements on their respective states in each round, but for now we do not specify whether there are memory effects (in the sense of Process~\ref{proc2}) across the rounds. In that case, we could consider the following contrived protocol:
\begin{algorithm}[htp]
\caption*{\textbf{Example protocol}}
\begin{algorithmic}[1]
\State For each round $j$ (starting from $j=1$), the following occurs:

\begin{itemize}
\item If $j$ is odd, Alice and Bob each choose (uniformly at random) either an $X$ or $Z$ measurement on their share of the state, and store the outcomes.

\item If $j$ is even, Alice and Bob each perform the same measurement as they did in the preceding round, and \emph{publicly announce} the outcomes.
\end{itemize}

\State Alice and Bob now completely ignore all the data from the even-numbered rounds, and with the remaining data (i.e.~from the odd-numbered rounds), perform all the classical postprocessing steps of a ``standard'' entanglement-based BB84 protocol, i.e.~basis announcements, sifting, parameter estimation, and so on. 

\end{algorithmic}
\end{algorithm}

It is straightforward to argue that this protocol is secure in a Process~\ref{proc1} model --- first observe that in that scenario, the measurements on different rounds commute since they act on separate tensor factors, hence we can take all the measurements in the odd-numbered rounds to be performed first, deferring the measurements on the even-numbered rounds (and corresponding announcements) until the very end of the following analysis. In that case, denoting the joint pre-measurement state (for Alice and Bob) for all the odd-numbered rounds as 
$\rho_\mathrm{odd}$, 
we trivially have that (before the measurements are performed) the quantum registers in the even-numbered rounds are just part of an extension of $\rho_\mathrm{odd}$. Observe that performing the above example protocol with the steps in even-numbered rounds deferred (i.e.~just measuring $\rho_\mathrm{odd}$ and performing classical postprocessing steps) corresponds exactly to performing a ``standard'' entanglement-based BB84 protocol on $\rho_\mathrm{odd}$. Therefore, it would be secure against any extension of this state that Eve might hold; in particular, the extension that includes the quantum registers from the even-numbered rounds. To finish up\footnote{This last step is to resolve the issue that the preceding analysis technically only proves security for a state where the even-numbered registers are still ``unmeasured'', whereas in the example protocol the measurements in even-numbered rounds depend on the measurements in odd-numbered rounds, which could potentially introduce additional correlations.} the argument, we simply need to observe that in a ``standard'' entanglement-based BB84 protocol, the parties announce their basis choices after the measurements have been performed. Therefore, these basis choices are included in Eve's side-information at the end of the process we have just described, and this is enough for her to reproduce the actual final state produced by the above example protocol (which also implies its security, by the data-processing inequality for trace distance).

On the other hand, there is a Process~\ref{proc2} model under which this protocol is trivially insecure --- namely, suppose that in the odd-numbered rounds, the devices share and measure a perfect Bell state $\ket{\Phi^+}$, but in the even-numbered rounds, the devices instead measure the post-measurement state from the preceding round. In that case, with the way the measurements are chosen in this protocol, the output in every even-numbered round is a perfect copy of the output in the preceding round. Since the even-numbered outputs are publicly announced, this makes the protocol completely insecure, as Eve learns \emph{all} the outputs that Alice and Bob obtained.

With this example, we can also gain some insight into a specific important security difference between Process~\ref{proc1} and Process~\ref{proc2} models. Namely, in a Process~\ref{proc1} model, suppose that we somehow had a guarantee that in some round $j$, the devices were genuinely measuring a Bell state $\ket{\Phi^+}_{\qA_j \qB_j}$. In that case, we would know that regardless of the structure of the pre-measurement quantum state across all the rounds, the $j^\text{th}$-round outputs at least are perfectly independent of outputs in all the other rounds (as well as any quantum side-information Eve has). This is an immediate consequence of the fact that $\ket{\Phi^+}_{\qA_j \qB_j}$ is already pure, and hence must be in product with any extension, in particular the parts of the state in other rounds and Eve's side-information. In sharp contrast, however, such a guarantee is markedly insufficient to draw the same conclusion in a Process~\ref{proc2} model --- as in the example above, the outputs in later rounds could potentially be exact copies of the $j^\text{th}$-round outputs. This means that not only do such outputs contribute no additional entropy, they can even leak information about outputs that would otherwise have been perfectly secret. (A ``purity'' argument of this form is often implicitly used in the proof of cPA theorems, and hence it seems likely that special care would be needed to handle this issue when trying to extend such results to Process~\ref{proc2} models.)

We also note that recently, a non-IID memory-based attack (i.e.~under a Process~\ref{proc2} model) was found~\cite{arx_SW23} on a DIQKD protocol that involves some public announcements depending on the parties' outputs (this attack has similarities to that presented in a previous work~\cite{TTB+16} on a related protocol). Qualitatively, the attacks in those works seem related to the form of behaviour presented above --- informally, the idea in those attacks is to prepare a ``perfect'' state in some of the rounds (hence generating a perfectly secret bit when considering that round in isolation), but then store the value of the secret bit in memory and leak it via the public announcements made in later rounds.
Hence this seems to again reflect the issue that in a Process~\ref{proc2} model, the parties' outputs in later rounds can be correlated ``more strongly'' with previous rounds as compared to Process~\ref{proc1} models, and hence can leak more information about previous rounds' outputs, even if those outputs had been ``perfectly secure'' at the time of generation. 

Another noteworthy point regarding the~\cite{TTB+16,arx_SW23} attacks is
that if we had supposed the quantum systems in those DIQKD protocols were described by a Process~\ref{proc1} model and had some large but finite dimension, it would have been possible to use a dF argument to show that those protocols are asymptotically secure (albeit with very slow convergence if the dimension is very large), thus ruling out the possibility of implementing these attacks (or anything similar) in a Process~\ref{proc1} model. Hence these attacks again emphasize the point that there can be protocols where there exist security proofs under a Process~\ref{proc1} model, but the protocol becomes insecure under a Process~\ref{proc2} model.

\begin{remark}
It is perhaps worth noting that the idea behind this form of attack may be of a different nature than the issues raised in the preceding sections. Specifically, the attacks here basically rely on ``reading'' the outputs of previous rounds from memory and then explicitly leaking them, whereas the issues raised in previous sections are more regarding the inability to connect the observed test-measurement outcomes to the actual quantities required for the security proof. It seems not entirely clear whether these can be viewed as manifestations of the same underlying issue.
\end{remark}

\section{Applicable proof techniques}
\label{sec:valid}

The more recently developed entropy accumulation theorem (EAT) and quantum probability estimation (QPE) frameworks are essentially developed precisely for a situation where sequential behaviour occurs, and hence can handle Process~\ref{proc2} models fairly naturally (as long as a particular Markov condition holds in the protocol). 
Furthermore, in the DI case there were also earlier security proof techniques for sequential behaviour~\cite{PM13,NBS+18,VV14}, as well as proofs based on parallel repetition theorems~\cite{JMS20,arx_Vid17}, which can also handle this scenario (up to a minor issue regarding whether the devices can signal inputs to each other after each round), albeit at the cost of lower asymptotic keyrates.
The fact that these proof techniques exist at least shows that most existing QKD protocols should not be entirely insecure in Process~\ref{proc2} models, as long as they can be analyzed in a manner compatible with use of these theorems. Somewhat speculatively, our discussions above also perhaps suggest why certain features of these theorems, which might appear peculiar at first glance, may in fact be necessary to handle Process~\ref{proc2} models. (In particular, the contrived protocol in the previous section seems very difficult to cast within the EAT/QPE frameworks --- this is consistent with the fact that it is genuinely insecure in a Process~\ref{proc2} scenario.)

For instance, protocols based on the EAT/QPE frameworks often use a ``spot-checking'' approach for choosing the test rounds, rather than a fixed-size random-sampling approach. It seems possible that this is because for the former, 
one has some notion of a similar ``structure'' in each round that allows a somewhat individual analysis of each round, even in a Process~\ref{proc2} model --- roughly, a test/generation decision is made independently in each round, followed by some measurement.
In contrast, for the latter, one cannot easily focus on an individual round in the analysis, because the test/generation decision is not independent of the other rounds. This is fine when analyzing a Process~\ref{proc1} model where the NS structure allows one to construct the ``counterfactual'' arguments on the global state that are used in the cPA/EUR techniques, but seems harder to use for Process~\ref{proc2} models.

Another point of interest may be the Markov condition in the EAT, which seems related to the observation in Sec.~\ref{sec:attacks} that in Process~\ref{proc2} models, later rounds have much more ``power'' (as compared to Process~\ref{proc1} models) to leak information about earlier rounds. Although, in the context of the EAT-based DIQKD security proofs in~\cite{ARV19,arx_TSB+20}, there are in fact two separate aspects of such leakage, only one of which is related to the Markov condition. Namely, the public announcement of the \emph{inputs} is indeed handled by noting that they satisfy the Markov condition, which fully prevents them from leaking information about earlier rounds. However, there is also a public announcement of test-round \emph{outputs}\footnote{Or if the test-round outputs are compressed into an error-correction string, this slightly increases the length of the error-correction string, which is also compensated for by subtracting this extra length from the smoothed min-entropy using a chain rule.} --- this is not handled within the EAT itself, but instead simply by using a coarse chain rule to bound the decrease in smoothed min-entropy in terms of the number of communicated bits (this coarse approach restricts us to only considering protocols where the test-round fraction is fairly small). 
A recently developed generalization of the EAT~\cite{arx_MFSR22} allows for directly conditioning on the test-round outputs being publicly announced (in the case of device-dependent QKD), but implicitly relies on a no-memory condition regarding how those outputs are produced. It remains to be seen whether further relaxations of the theorem conditions are possible, but it would be important to keep in mind that any such relaxation could only be to an extent consistent with the existence of the attacks described in this work.

\printbibliography

\end{document}